\documentclass[twocolumn]{pasj00}

\begin{document}
\SetRunningHead{Takami et al.}{SiO $J$=8-7 Emission towards HH 212}
\Received{2006/01/??}
\Accepted{2006/??/??}

\title{Kinematics of SiO $J$=8-7 Emission towards the HH 212 Jet}

\author{Michihiro \textsc{Takami}\altaffilmark{1},
        Shigehisa \textsc{Takakuwa}\altaffilmark{2},
        Munetake \textsc{Momose}\altaffilmark{3},
        Masa \textsc{Hayashi}\altaffilmark{1}, \\
        Christopher J. \textsc{Davis}\altaffilmark{4},
        Tae-Soo \textsc{Pyo}\altaffilmark{1},
        Takayuki \textsc{Nishikawa}\altaffilmark{1},
        and
        Kotaro \textsc{Kohno},\altaffilmark{5}
} \altaffiltext{1}{Subaru Telescope, 650 North A'ohoku Place, Hilo,
Hawaii 96720, USA} \altaffiltext{2}{National Astronomical
Observatory of Japan, Osawa, Mitaka, Tokyo 181-8588, Japan}
\altaffiltext{3}{Institute of Astronomy and Planetary Science,
Ibaraki University, Bunkyo 2-1-1, Mito, Ibaraki 310-8512, Japan}
\altaffiltext{4}{Joint Astronomy Centre, 660 North A'ohoku Place,
University Park, Hilo, Hawaii 96720, USA} \altaffiltext{5}{
Institute of Astronomy, The University of Tokyo, 2-21-1 Osawa,
Mitaka, Tokyo 181-0015}


%


\maketitle

\begin{abstract}
We present SiO $J$=8-7 (347.3 GHz) observations towards HH 212 using
the ASTE telescope. Our observations with a
22''-diameter beam show that the SiO emission is highly concentrated
within 1' of the driving source. We carefully compare the SiO
observations with archival H$_2$ 1-0 S(1) images and published H$_2$
echelle spectra. We find that, although the SiO velocities closely
match the radial velocities seen in H$_2$, the distribution of H$_2$
and SiO emission differ markedly. We attribute the latter to the
different excitation conditions required for H$_2$ and SiO emission,
particularly the higher critical density ($n_{H_2} \sim 10^8$
cm$^{-3}$) of the SiO $J$=8-7 emission. The kinematic similarities
imply that the H$_2$ and SiO are associated with the same internal
working surfaces. We conclude that the SiO $J$=8-7 emission has a
potential to probe the jet/wind launching region through
interferometric observations in the future, particularly for the
youngest, most deeply embedded protostars where IR observations are
not possible.

\end{abstract}

\section{Introduction}
Jets/outflows are a key phenomenon for the evolution of young
stellar objects (YSOs). Theories predict that these outflows remove
excess angular momentum from the accreting material, thereby
allowing mass accretion to occur (see e.g., Shu et al. 2000;
K\"onigl et al. 2000). Observing the kinematics in the flow
acceleration region is important if we are to understand this
physical link. Unfortunately, there are two crucial problems which
hinder the investigation through this approach. One is the limited
spatial resolution of telescopes: these are in most cases
insufficient to fully resolve the jet/wind launching region, which
is believed to be located within a few AU of the driving source. The
other is the fact that such a region is often deeply embedded within
a massive envelope, making it difficult to observe at optical to
near-IR wavelengths.

The rotation of jets/winds is arguably easier to observe than the
kinematics in their launching regions. Such observations are
therefore useful for the understanding the mechanism of mass
ejection and accretion. Several observations have been attempted at
optical-IR wavelengths, providing results consistent with the
magneto-centrifugal wind scenario (see, e.g., Davis et al. 2000;
Bacciotti et al. 2002; Coffey et al. 2004; Woitas et al. 2004).
However, this approach may be hampered by spurious wavelength shifts
of the grating spectrographs, or the interaction of the jet with the
inhomogeneous ambient medium (cf. Bacciotti et al. 2002).

Millimeter/sub-millimeter interferometry in the near future has a great
potential to overcome these difficulties for studying kinematics in
the jet/wind launching region, and also measuring flow rotation.
We have therefore begun to search for appropriate emission lines for
these studies using the ASTE (Atacama Submilimeter Telescope
Experiment) telescope, a single-dish 10-m submillimeter telescope in
northern Chile. ASTE is a pioneering activity on submillimeter
astronomy at the ALMA site (Ezawa et al. 2004). In this paper we
present SiO $J$=8-7 (347.3 GHz) observations towards HH 212. SiO
emission at mm-submm wavelengths has been observed towards a number
of outflows associated with low-mass YSOs (see e.g., Mikami et al.
1992; Codella, Bachiller \& Reipurth 1999; Hirano et al. 2001;
Nisini et al. 2002, 2005; Gibb et al. 2004). The emission associated
with these objects is likely to originate in shocks via grain
sputtering or grain-grain collisions releasing Si-bearing material
into the gas phase (e.g., Schilke et al. 1997; Caselli, Harquist \&
Havenes 1997; Pineau des For\^ets et al. 1997). The observations to
date have revealed that the kinematics in SiO emission markedly
differ from millimeter CO emission, a well-known probe of
molecular bipolar outflows (e.g., Hirano et al. 2006; Palau et al.
2006).
HH 212 is a highly symmetric two-sided jet in the Orion molecular
cloud (Zinnecker, McCaughrean \& Rayner 1998), lying within
5$^\circ$ of the plane of the sky (Claussen et al. 1998).
Interferometric observations of millimeter CO/SO emission have been
conducted by Lee et al. (2000, 2006) with an angular resolution down
to a few arcsecond, revealing the origins of these lines.

In this paper we show that the SiO $J$=8-7 line in this object is
probably associated with the ejecta of the collimated jet, in
particular its internal working surfaces. Due to this fact, together
with its high critical density ($n_{H_2} \sim 10^8$ cm$^{-3}$), this
line has a potential to probe the jet/wind launching region through
interferometric observations in the future.

\section{Observations and Results}
Observations were made using the ASTE telescope on August 27--28 2005.
The antenna
diameter of 10-m provides a FWHM beam size of 22" at 350 GHz,
corresponding to 0.05 pc at a distance of 450 pc. The telescope
is equipped with an SIS heterodyne-receiver and an XF-type digital
auto-correlator that gives a 0.5 MHz frequency resolution over the total
bandwidth of 512 MHz. The weather was quite good, with a system temperature of
180--260 K. The pointing was checked every few hours and was found
to be accurate to better than 3". OMC-1 was also observed for
calibration with standard spectra. Performing comparisons with the
data obtained at the CSO 10.4-m telescope (Schilke et al. 1997), we derive the
main beam efficiency $\eta_{MB}$=0.73--0.74 during the observations.

The spectra were obtained towards three well-known low-mass outflow
sources, including a Class 0 protostar (HH 212) and two Class I
protostars (SVS 13 and L1551-IRS 5), with an initial resolution of
0.5 MHz. All of these objects exhibit bright H$_2$ emission due to
shocks (e.g., Davis et al. 1995, 2000, 2002, 2006). In HH 212 and
SVS 13, seven and two points were observed along the jet,
respectively. Each spectrum was binned to increase signal-to-noise,
providing a velocity resolution of 4.3 km~s$^{-1}$. The results are
summarized in Table \ref{flux}. We detect the emission clearly at
the driving source and 20'' north/south, and marginally at 40''
north/south in HH 212. Upper limits to the brightness temperature of
0.02--0.03 K are obtained for the remaining two positions and for the
other two targets. The non-detection towards the two Class I protstars
corroborate previous SiO $J$=5-4 observations, which show that
SiO emission is preferentially detected towards Class 0
protostars (Gibb et al. 2004).


\begin{table*}
  \caption{SiO $J$=8-7 Brightness Temperatures and Integrated Intensities}\label{flux}
  \begin{center}
    \begin{tabular}{lcccccc} \hline \hline
Object & R.A. (2000) & Dec. (2000) &
Jet P.A.\footnotemark[a] & Offset &
Peak $T_{MB}$ & $\int T_{MB}~dv$ \\
&&& (deg.) & (arcsec) & (K) &
(K~km~s$^{-1}$) \\ \hline
HH 212 & 5$^h$43$^m$51.0$^s$ & --01$^\circ$02'52"   & 22  & +60  & $<$0.02       & --- \\
       &                     &                      &     & +40  & 0.06$\pm$0.02 & ---\footnotemark[b] \\
       &                     &                      &     & +20  & 0.22$\pm$0.03 & 2.98$\pm$0.07\\
       &                     &                      &     &  0   & 0.19$\pm$0.02 & 3.81$\pm$0.04\\
       &                     &                      &     & --20 & 0.21$\pm$0.03 & 2.26$\pm$0.04\\
       &                     &                      &     & --40 & 0.06$\pm$0.02 & ---\footnotemark[b] \\
       &                     &                      &     & --60 &$ <$0.03       & --- \\
SVS 13 & 3$^h$29$^m$03.7$^s$ & +31$^\circ$16'04"    & 123 & 0     & $<$0.02       & --- \\
       &                     &                      &     & +20   & $<$0.02       & --- \\
L1551-IRS 5 & 4$^h$31$^m$31.3$^s$ & +18$^\circ$08'05"& ---& 0     &
$<$0.02       & --- \\ \hline \hline
\multicolumn{7}{@{}l@{}}{\hbox to 0pt{\parbox{145mm}{\footnotesize
\footnotemark[a]{Davis et al. (2000)}
\par\noindent
\footnotemark[b]{We cannot measure the integrated intensity since the signal-to-noise is not
sufficient for determining the velocity range.}
     }\hss}}
\end{tabular}
\end{center}
\end{table*}

\begin{figure*}
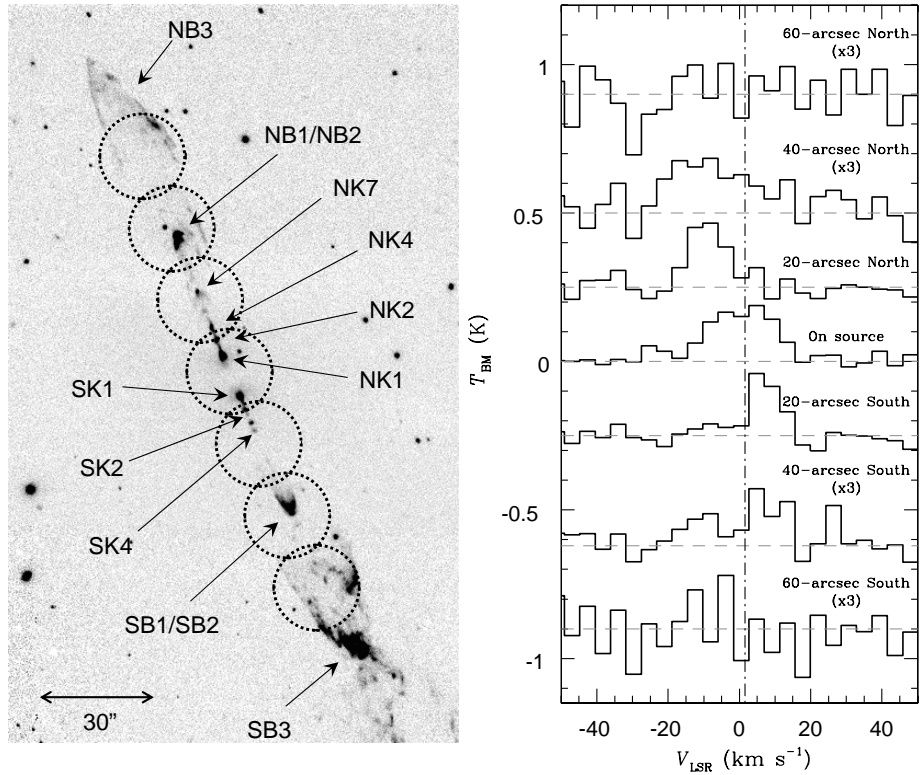

\begin{center}
\FigureFile(60mm,50mm){f1l.epsi}
\FigureFile(60mm,50mm){f1r.epsi}
\end{center}
\caption{ (left) Narrow-band H$_2$ 1-0 S(1) (2.122$\micron$) image
of HH 212 obtained using VLT-ISAAC. The raw images were obtained
from the ESO Archive. The name of the knots are based on Zinnecker
et al. (1998). Dotted circles indicate the positions and beam size
(22'') of the ASTE observations. (right) SiO $J$=8-7 line profiles
observed at seven positions. The brightness temperature at 40--60''
north/south is multiplied by 3 to emphasize marginal features. The
dot-dashed line indicates the systemic velocity of the molecular
core ($V_{LSR}$=1.6 km~s$^{-1}$,
 Wiseman et al. 2001)
}
  \label{f1}
\end{figure*}

Figure \ref{f1} shows the position and line profiles observed
towards HH 212.
The line profile at 20'' north shows only a
blueshifted component, while that at 20'' south shows only a
redshifted component.
The line profiles observed at 20'' north and south peak at
--8 and 5 km~s$^{-1}$, respectively: i.e., $\sim$10 km~s$^{-1}$
blueshifed and $\sim$3 km~s$^{-1}$ redshifted with respect to the systemic
velocity of the cloud core (1.6 km~s$^{-1}$, Wiseman et al. 2001).
These flow velocities are markedly larger than those observed in CO $J$=1-0
emission, whose peaks are only 1-2 km~s$^{-1}$
blueshifted/redshifted (see Lee et al. 2000).


\begin{figure}
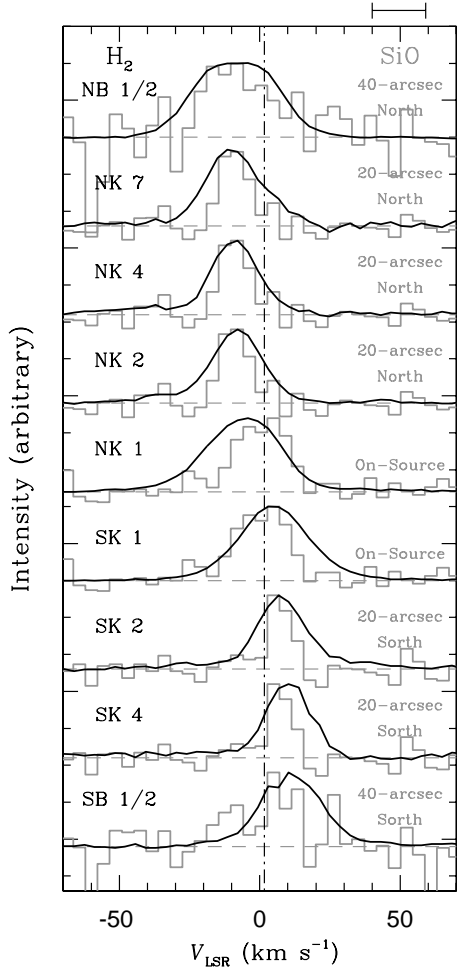

\begin{center}
\FigureFile(60mm,50mm){HH212_prof.epsi}
\end{center}
\caption{
H$_2$ 1-0 S(1) line profiles observed at each knot (solid),
together with the SiO $J$=8-7 profiles (histogram). Each knot is covered by
the SiO beam as follows: NB 1/2 by the 40'' north beam; NK 2--7 by the 20'' north beam;
NK 1 and SK 1 by the on-source beam; SK 2--4 by  the 20'' south beam; and SB 1/2 by
the 40'' south beam.
The bar above the
figure shows the FWHM of the instrumental broadening for the H$_2$
emission. The dot-dashed line indicates the systemic velocity. }
  \label{H2_profs}
\end{figure}

Davis et al. (2000) obtained three adjacent, parallel long-slit spectra of H$_2$ 1-0
S(1) emission (2.122 $\micron$) along the axis of the HH 212 jet,
with a 0.5'' separation between the slits. To perform comparisons of kinematics between
the H$_2$ and SiO emission, we summed these three spectral images and plot H$_2$ profiles
in Figure \ref{H2_profs} for individual emission knots. The SiO
profiles at the adjacent positions are also shown. The figure shows
that the velocity observed in the SiO emission is very similar to that of
the H$_2$ emission. Indeed, the northern H$_2$ knots and bow shock, NK
1--7 and NB 1/2, exhibit a peak velocity of --4 to --12 km~s$^{-1}$,
matching well with the SiO peak velocity observed at 20'' north of
the driving source (--8 km~s$^{-1}$). The southern knots and bow
shock (SK 1--4 and SB 1/2) exhibit a peak velocity of 5--10
km~s$^{-1}$. The H$_2$ peak velocity at SK 1 coincides with the SiO
peak velocity, whereas the H$_2$ peak velocity observed in the other
structures are slightly larger than the SiO peak velocity. The H$_2$
profiles at NK 2--7 and SK 3--4 shows FWHM velocities of 19--23
km~s$^{-1}$, significantly larger than the SiO emission at the
corresponding beam (10--15 km~s$^{-1}$ at 20'' north/south). Note, however,
that this is due to the instrumental broadening of the H$_2$
profiles (19 km~s$^{-1}$). We derive the deconvolved FWHM velocities to be less than
15~km~s$^{-1}$ for NK 2--7 and SK 3--4 assuming $v_{obs}^2=v_{dec}^2+v_{ins}^2$, where
$v_{obs}$, $v_{dec}$ and $v_{ins}$ are the FHWM velocity of the observed, deconvolved and instrumental
profiles, respectively.  The H$_2$ 1-0 S(1) peak and FWHM
velocities for the individual knots are tabulated in Table
\ref{H2param}


\begin{table}
\begin{center}
\caption{H$_2$ 1-0 S(1) Peak and FWHM velocities}\label{H2param}
\begin{tabular}{lrrcc} \hline \hline
Knot & Offset\footnotemark[a] &
Peak $V_{LSR}$\footnotemark[b] & \multicolumn{2}{c}{$V_{FWHM}$}\\
&&& observed & deconvolved \footnotemark[c]\\
& (arcsec) & (km~s$^{-1}$) & (km~s$^{-1}$) & (km~s$^{-1}$) \\ \hline
NB 1/2  & 39 & --12  & 31 & 24\\
NK 7    & 24 & --8   & 21 & ~9\\
NK 4    & 13 & --8   & 20 & ~6\\
NK 2    & 10 & --8   & 23 & 12\\
NK 1    & 5& --4     & 36 & 31\\
SK 1    & --6& 5     & 28 & 21\\
SK 2    & --10& 7    & 20 & ~6\\
SK 4    & --14& 10   & 19 & ---\\
SB 1/2  & --40& 10   & 25 & 16\\ \hline \hline
\multicolumn{4}{@{}l@{}}{\hbox to 0pt{\parbox{88mm}{\footnotesize
\footnotemark[a]{Based on Zinnecker et al. (1998). }
\par\noindent
\footnotemark[b]{The uncertainty is $\pm$2 km~s$^{-1}$.}
\par\noindent
\footnotemark[c]{Assuming $v_{obs}^2=v_{dec}^2+v_{ins}^2$, where
$v_{obs}$, $v_{dec}$ and $v_{ins}$ are the FHWM velocity of the observed, deconvolved and instrumental
profiles, respectively.}
    }\hss}}
\end{tabular}
\end{center}
\end{table}

\begin{figure}
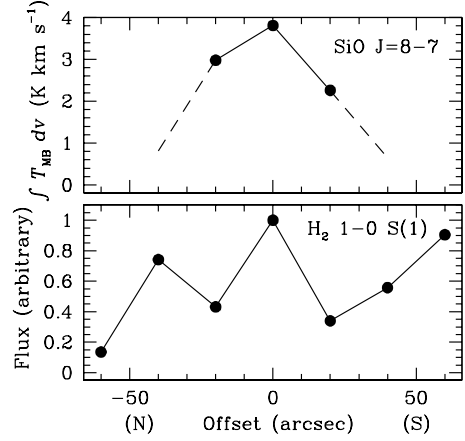

\begin{center}
\FigureFile(60mm,50mm){SiO_H2.epsi}
\end{center}
\caption{
Intensity distribution of SiO $J$=8-7 and H$_2$ 1-0 S(1) emission measured using
the same size apertures. Uncertainties are comparable or smaller than the
filled circles. }
  \label{intensity}
\end{figure}

Although the H$_2$ and SiO velocities are closely correlated, their disributions
differ markedly. The SiO $J$=8-7 flux is highly concentrated within 1' of the driving
source. This contrasts with the distribution of the H$_2$ 1-0 S(1)
and CO $J$=1-0 emission, which extend over a minute along the jet axis
(see Figure \ref{f1}; Lee et al. 2000).
The strongest SiO emission coincides with regions where H$_2$ knots are relatively
faint (20'' north and south of the source), while the brightest H$_2$ knots NB 1/2 and SB 1/2 correspond
with weaker SiO emission.
To show this quantitatively,
we reduced narrow-band H$_2$ images obtained using VLT-ISAAC on
October 19 and November 7 2003, and analyzed the images as follows. After
sky-subtraction and flat-fielding, the flux from the
foreground/backgound stars were carefully removed, then the H$_2$
flux above the 5-sigma level (corresponding to 2\% of the peak flux
at SK 2) was measured in each gaussian beam with a FWHM of 22''.
Figure \ref{intensity} shows the intensity distribution of H$_2$
emission obtained in this way. The distribution of the SiO $J$=8-7
integrated intensity is also shown. The H$_2$ flux distribution
shows flux minima at 20'' north and south, and the flux is greater
at 40'' north and south by a factor of $\sim$2. The H$_2$ flux at
60'' south is comparable to that at the driving source. In contrast,
the SiO emission is detected at 40'' north/south only marginally
($\sim$3-$\sigma$), and not detected at 60'' north/south.
Although the integrated intensity at these regions are uncertain, due to
the uncertainty of the line profile, the brightness temperature suggests that the
intensity at 40''--60'' north/south is smaller than the inner region
by a factor of at least 3 \footnote{0.8$\pm$0.3 and $<$0.3 K km s$^{-1}$ for 40'' and 60''
north, respectively, assuming the same line profile as 20'' north;
0.6$\pm$0.2 and $<$0.3 K km s$^{-1}$ for 40'' and 60''
south, respectively, assuming the same line profile as 20'' south.}.


\begin{figure}
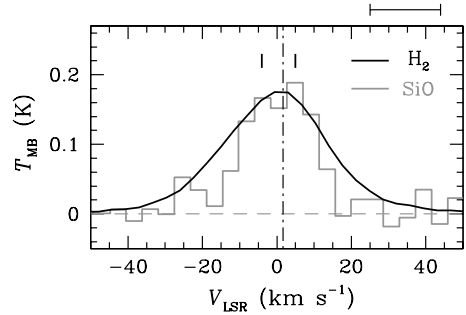

\begin{center}
\FigureFile(60mm,50mm){f4new.epsi}
\end{center}
\caption{
The H$_2$ 1-0 S(1) profile averaged from NK 1 and SK 1 (solid), and
the SiO $J$=8-7 profile at the driving source (historgram).
The marks at the top of the line profiles
indicate the peak velocity for NK 1 and SK 1.
The dot-dashed line indicates the
systemic velocity of the molecular core. The bar at the top-right shows the
FHWM of the instrument broadening for the H$_2$ profile.}
  \label{SiO_middle}
\end{figure}

In Figure \ref{H2_profs}, the SiO line profile at the driving source
shows flow velocities slower than the H$_2$ emission at NK 1 and SK
1, the brightest H$_2$ features in the beam. To investigate this in detail,
we averaged the H$_2$ profiles from NK 1 and SK 1,
and plot the results in Figure \ref{SiO_middle}
together with the SiO profile at the driving source. In Figure \ref{SiO_middle}
the velocity centoid is located at the systemic velocity for both SiO and H$_2$ emission.
The SiO profile exhibits two marginal peaks at positive and negative velocities, and these
match well with the peak velocity of NK 1 and SK 1, respectively.
The FWHM of the H$_2$ profile is 32~km~s$^{-1}$, broader than SiO ($\sim 20$~km~s$^{-1}$),
although again the difference is attributed to the instrumental broadening of the former profile.
We thus conclude that the SiO emission observed towards the driving source is associated
with these H$_2$ knots.


Gibb et al. (2004) observed the SiO $J$=5-4 emission (217.1 GHz) at
the driving source and 20'' south using the JCMT, which provides the
same beam size as the SiO $J$=8-7 observations at ASTE. The $J$=5-4
profile observed at the driving source is markedly different from
the $J$=8-7 profile: the JCMT observations show a blueshifted component
peaking at --6 km~s$^{-1}$, and a possible redshifted component
peaking close to 28 km~s$^{-1}$. However, the $J$=5-4 and
$J$=8-7 line profiles do resemble each other at 20'' south. Using the
$J$=5-4 integrated intensity of 2($\pm$0.6) K~km~s$^{-1}$, we derive
the 5-4/8-7 intensity ratio of $\sim$1. Performing comparisons with
model calculations by Nisini et al. (2002), we derive a molecular
hydrogen number density of 10$^{6}$--10$^{7}$ cm$^{-3}$ assuming a
temperature $T$=100--1000 K. This value is 10--100 times larger than
that obtained by Gibb et al. (2004) using the 2-1/5-4 intensity
ratio ($n_{H_2} \sim 10^5$ cm$^{-3}$). The discrepancies between our
results and those of Gibb et al. (2004) would be attributed to the fact that
the upper transitions have higher critical densities ($1 \times 10^6$,
$2 \times 10^7$ and $1 \times 10^8$ cm$^{-3}$ at $T$=100 K for the $J$=2-1, 5-4
and 8-7 transitions, respectively).

\section{Discussion}
Millimeter/sub-millimeter SiO emission has been observed towards a
number of molecular outflows for the last decade. Studies to date
have shown that the emission originates from the regions with a
temperature $T > 100$ K and a molecular hydrogen density $n_{H_2}
> 10^5$ cm$^{-3}$ (see e.g., Codella et al. 1999; Nisini et al.
2002, 2005; Gibb et al. 2004), both much higher than the regions
associated with mm CO emission. The abundance of gas-phase SiO in
these regions is estimated to be $10^{-11}$ to $10^{-6}$,
significantly higher than that of quiescent clouds ($<$10$^{-12}$,
Ziurys, Friberg \& Irvine 1989). This supports the theories, which
predict that shocks release silicon-bearing molecules from dust to
the gas, allowing bright SiO emission to be observed towards
molecular outflows (Schilke et al. 1997; Caselli et al. 1997; Pineau
des For\^ets et al. 1997).

Chandler \& Richer (2001), Hirano et al. (2006) and Palau et al.
(2006) conducted high-resolution observations of SiO $J$=1-0, 5-4
and 8-7 lines, respectively, towards HH 211. Performing comparisons
with the H$_2$ 1-0 S(1) image, theses authors show that the SiO
emission is associated with the collimated jet. Our results for HH
212 show that the kinematics of the SiO emission closely resemble
the near-infrared H$_2$ emission associated with the collimated jet,
agreeing with this picture. Since the SiO emission shows nearly the
same velocity as the H$_2$ emission, it is likely that the emission
is associated with the ejecta of the collimated jet, in particular
with its internal working surfaces. This contrasts with millimeter
CO emission, which is markedly slower than the SiO and H$_2$
emission and probably originates from gas entrained by the jet (Lee
et al. 2000), or limb-brightened shells surrounding the H$_2$ jet
(Lee et al. 2006).


While the observed kinematics of the SiO $J$=8-7 emission match
well  with the H$_2$ 1-0 S(1) emission, their intensity
distributions markedly differ with each other due to different
excitation conditions. Indeed, the shock-excited H$_2$ 1-0 S(1)
emission originates from regions with a temperature of $\sim 2
\times 10^3$ K (see, e.g., Eisl\"offel et al. 2000; Takami et al.
2006), while the SiO emission presumably originates from regions with
a much lower temperature due to its low upper energy level ($E_u/k =
75$ K for $J$=8-7). In addition, the critical density of H$_2$ 1-0
S(1) is $\sim 10^6$ cm$^{-3}$ for H$_2$--H$_2$ collision, while that
of SiO $J$=8-7 is $\sim$100 times larger ($1 \times 10^8$
cm$^{-3}$). The enhancement of the $J$=8-7 emission within 1' of the
driving source is attributed to the density gradient in the jet. The
same trend of the SiO $J$=8-7 intensity distribution is also
observed in HH 211 by Palau et al. (2006): these authors measured
5-4/8-7 ratios along the jet, revealing that the gas density is
higher near the driving source.

Due to its high critical density, the SiO $J$=8-7 line has the
potential to probe the jet/wind launching regions through
interferometric observations in the future. However, the emission
has not been detected even towards SVS 13 and L1551-IRS~5, which
exhibit a bright molecular-hydrogen-emission-line (MHEL) region
within a few second of the driving source (Davis et al. 2001, 2002,
2006; Takami et al. 2006). This could be due to the small angular
scale of the emission region, and so beam dilution effects, or to the
high temperature. Alternatively, the gas density and/or SiO abundance
may be too low in there less embedded, more evolved Class I sources.
Observations with
a higher angular resolution or better signal-to-noise would allow
for investigating this in detail.
\vspace{20pt}

We acknowledge the ASTE team for their excellent support. We thank
Dr. Hirano for useful discussion. The narrow-band images of the
H$_2$ 1-0 S(1) emission have been obtained through the ESO archive
operated by European Southern Observatory. This research has been
made use of the Simbad database operated at CDS, Strasbourg, France,
and the NASA's Astrophysics Data System Abstract Service.



\end{document}